\documentclass[12pt]{iopart}
\usepackage{amsfonts}
\usepackage{amssymb}
\usepackage{graphicx}
\usepackage{bbold}
\usepackage{textcomp}

\begin{document}

\title[Engineering integrated pure narrow-band photon sources]{Engineering integrated pure narrow-band photon sources}

\author{E.~Pomarico$^{1}$, B.~Sanguinetti$^{1}$, C.~I.~Osorio$^{1}$, H.~Herrmann$^{2}$,  R.~T.~Thew$^{1}$.}

\address{$^{1}$Group of Applied Physics, University of Geneva, 1211 Geneva,
Switzerland.}
\address{$^{2}$Angewandte Physik, University of Paderborn, 33095
Paderborn, Germany.} \ead{enrico.pomarico@unige.ch}

\begin{abstract}
Engineering and controlling well defined states of light for quantum information applications is of increasing importance as the complexity of quantum systems grows. For example, in quantum networks high multi-photon interference visibility requires properly devised single mode sources. In this paper we propose a spontaneous parametric down conversion source based on an integrated cavity-waveguide, where single narrow-band, possibly distinct, spectral modes for the idler and the signal fields can be generated.
This mode selection takes advantage of the {\it clustering} effect, due to the intrinsic dispersion of the nonlinear material. In combination with a CW laser and fast detection, our approach provides a means to engineer a source that can efficiently generate pure photons, without filtering, that is compatible with long distance quantum communication. Furthermore, it is extremely flexible and could easily be adapted to a wide variety of wavelengths and applications.

\end{abstract}

\newpage
\tableofcontents

\section{Introduction}

Engineering pure quantum states of light is becoming increasingly important for a multitude of applications associated with quantum information technologies, such as the implementation of quantum repeaters~\cite{Duan2001,Simon2007,Sangouard2011}, device independent quantum key distribution (QKD)~\cite{Acin2007}, entangled states for quantum metrology~\cite{Dowling2008,Afek2010}, or for optical quantum computation~\cite{Knill2001,Raussendorf2001}. This is especially relevant for the scalability of resources in complex quantum systems: for instance, for the realization of a quantum communication network, for which multiple independent systems need to be interfaced with high interference visibility, engineering suited photon pair sources is crucial.

Recently, we have seen a concerted effort to develop pure photon sources via spontaneous parametric downconversion (SPDC) or four wave mixing ~\cite{GarayPalmett2007, Branczyk2011,Soller2011}. However, these approaches usually generate relatively broadband photons. Long distance quantum communication~\cite{Gisin2007, Gisin2010a}, on the other hand, requires narrow band photons, either for coupling to quantum memories~\cite{Sangouard2011}, or simply for robustness against thermal dilation of optical fibers~\cite{Halder2008}.  Moreover, we need these sources to be efficient, by which we mean that both the spatial and spectral modes have to be efficiently selected. The former requires optimal coupling into a single mode optical fiber, whereas the latter implies spectral purity, which is generally achieved via filtering.

The problem with filtering is that it is inherently lossy: a Gaussian-shaped filter, even if perfectly transparent at its centre wavelength, will always introduce some loss in the wings resulting in a transmission of only $1/\sqrt{2}\approx 70\%$ of the light inside its own spectrum. Emerging applications like device independent QKD~\cite{Acin2007}, as well as more fundamental challenges like all-optical loop-hole free Bell tests~\cite{Kwiat1994}, cannot be realized even with this trivial loss mechanism. Sources compatible with long distance quantum communication have so far followed two paths: SPDC sources pumped in a CW mode with extremely lossy filtering ~\cite{Halder2007,Clausen2011}, or optical parametric oscillators (OPO)~\cite{Lu2000, Wang2004, NeergaardNielsen2007, Scholz2009}. The latter provide quite a good solution, since they naturally produce narrow band photons and avoid the gaussian loss, but are generally complex and require complicated and time consuming (off-line) stabilization routines.

In this paper we introduce an integrated OPO approach, that exploits a spectral clustering effect, related to the intrinsic dispersion of the non linear crystal and observed in~\cite{Pomarico09}, to suppress undesired modes and obviate the need for filtering. This integrated OPO requires only a simple (and continuous) temperature stabilization. In conjunction with detectors that have a fast temporal response~\cite{KorneevSNSPD04,Thew08a}, this approach provides a simple, flexible and efficient narrow band pure photon source. Firstly, we explain the concept of an integrated OPO photon pair source, what the clustering effect is and how it arises from the dispersion of the nonlinear crystal. We then compare our model system with some preliminary experimental results and provide easily adaptable guidelines for how to design such a system for a wide variety of wavelengths and applications. An example of such a source is also given. Finally, we discuss the potential of our approach in the context of engineering complex quantum communication architectures.

\section{Integrated cavity-waveguide SPDC source}
For practical applications in quantum information and communication, having an integrated device~\cite{OBrien2009,Smith2009,Sansoni2010} offers a number of advantages. For example, stabilizing a high finesse doubly resonant cavity can prove challenging. In a monolithic design, such as the one provided by an integrated cavity-waveguide, only the temperature of a single component needs to be stabilized. It is also by changing the temperature that the device is phase-matched for a variety of wavelengths. Another advantage of integration is that very short cavity lengths can be achieved, allowing one to increase the free spectral range (FSR), which will be important for what follows.
\begin{figure}[h]
\begin{center}
\includegraphics[width=0.6\textwidth]{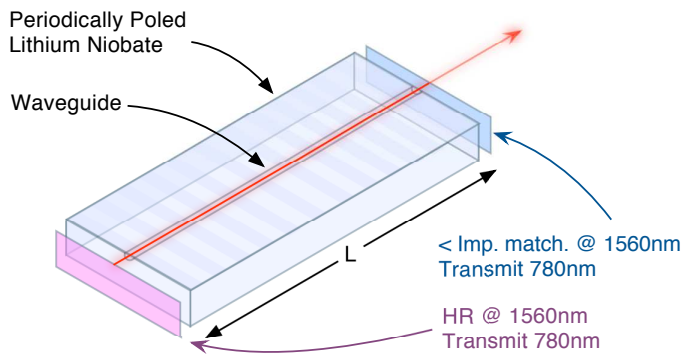}
\caption{Integrated cavity-waveguide formed by reflection coating mirrors at the entrance and exit surfaces of a periodically poled nonlinear material.}
\label{fig:waveguide}
\end{center}
\end{figure}

The generic structure of the proposed device is shown in figure~\ref{fig:waveguide} and consists of a periodically poled Lithium niobate (PPLN) crystal, in which a waveguide has been created~\cite{Chou1998,Tanzilli2001,Schreiber2001,Roussev04a}. Other periodically poled materials, such as PPKTP~\cite{Fiorentino2007}, can also be adopted. In an optimal device the entrance facet would be coated for high reflection at the signal and idler wavelength, whilst the exit facet would have an impedance-matched reflectivity sufficient to achieve the desired finesse, but low enough as to be the dominant loss mechanism.

We can treat this device as a Fabry-Perot cavity, whose finesse is completely determined by the losses and reflectivities:
\begin{equation}
\mathcal{F}=\pi^{-1} 2\arcsin\Big(\frac{1-\rho^{1/2}}{2 \rho^{1/4}}\Big),
\label{eq:finesse}
\end{equation}
where $\rho$ is related to the power in the cavity after a round trip and can be expressed as $\rho=R_1 R_2 10^{-2\alpha L/10}.$ $R_1$ and $R_2$ are the reflectivities of the mirrors, $\alpha$ is the absorption coefficient expressed in dB/cm, $L$ the length of the crystal in cm and $2$ accounts for the fact that the optical path per round trip is $2L$. In practice, the coating reflectivities $R_1$ and $R_2$ can be estimated by coating an optical blank with the same coating as the integrated device.  Once the finesse of the cavity is measured, one can solve equation~(\ref{eq:finesse}) for the waveguide loss coefficient $\alpha$. In our initial test device \cite{Pomarico09}, $\alpha$ was found to be 0.06\,dB/cm. Having low internal losses is important, as these will ultimately limit the achievable finesse for a given cavity length.

\subsection{Clustering in doubly-resonant SPDC}
\begin{figure}[t]
\begin{center}
\includegraphics[width=0.95\textwidth]{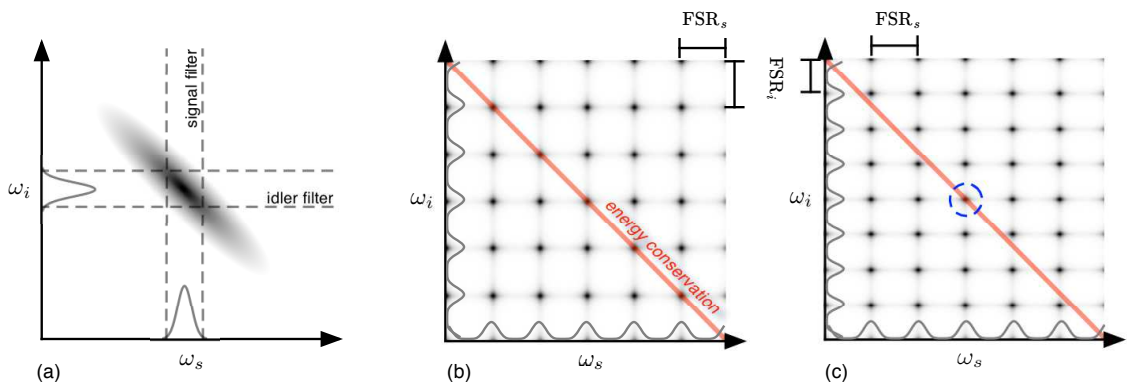}
\caption{The joint frequency spectrum showing: a) typical correlations between the signal and the idler photon frequencies; (b) correlations in a doubly-resonant SPDC source with no dispersion, where energy conservation (diagonal line) and resonance can be satisfied for a large number of adjacent cavity modes and (c) correlations when the medium is dispersive: the spacing of signal and idler modes is different and fewer modes satisfy the cavity and energy conservation constraints (one is shown in the dashed circle).}
\label{fig:joint_spectrum}
\end{center}
\end{figure}

The spectrum of photons generated by SPDC is usually determined by the phase-matching conditions between the pump, signal and idler fields in a nonlinear crystal~\cite{Fiorentino2007}, resulting in a typical joint frequency spectrum like that shown in figure~\ref{fig:joint_spectrum}a. When the nonlinear crystal is placed inside a cavity~\cite{Lu2000, Wang2004, NeergaardNielsen2007, Scholz2009}, an extra set of constraints must be satisfied: photons can only be emitted in modes which are supported by the resonator. For a Fabry-Perot cavity the resonant modes are given by the Airy function. This second constraint  on the joint frequency spectrum is represented in figure~\ref{fig:joint_spectrum}b. When pumping with a narrow CW laser, energy conservation ($\omega_s+\omega_i = \omega_p$) will further restrict all photons to be emitted on the diagonal of the joint spectral function (figure~\ref{fig:joint_spectrum}b). If the refractive index is equal for both signal and idler photons, then the optical path length of the cavity, and the Airy functions, will be the same for both photons such that the distance between the peaks will be defined by the FSR of the cavity.

\begin{figure}[t]
\begin{center}
\includegraphics[width=0.5\columnwidth]{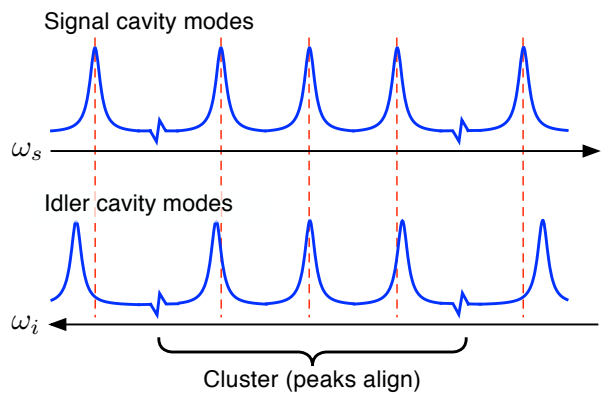}
\caption{In a dispersive medium the optical length of a cavity varies with wavelength. SPDC occurs in such a cavity only when the double resonance condition is satisfied, i.e. when both signal and idler frequencies (satisfying $\omega_p=\omega_s+\omega_i$) are resonant. }
\label{fig:FSR_2}
\end{center}
\end{figure}
However, if the signal and idler photons have different dispersion characteristics, for example when they are non-degenerate in frequency or have distinct polarizations, the FSR at the signal and idler wavelength will be different. In this situation, simultaneous resonance for both signal and idler arises only in regions of the spectrum where the peaks of the Fabry-Perot for the signal and idler align (figure~\ref{fig:joint_spectrum}c and figure~\ref{fig:FSR_2}). These regions are called "clusters"~\cite{Giordmaine1966,Eckardt1991}, as they usually contain a number of emission peaks. However, as we shall see below, it is possible to design the finesse of the cavity, so as to produce clusters containing only a single peak.

\subsection{Calculated and observed clustering spectrum} \label{par:clustering}

The state of the photon pairs emitted by a doubly resonant OPO is given by~\cite{Jeronimo-Moreno2010}:
\begin{equation}
|\Psi\rangle_{OPO}=\int d\omega_s \int d\omega_i S(\omega_s,\omega_i) a^{\dag}_s (\omega_s) a^{\dag}_i (\omega_i) |vac\rangle,
\end{equation}
where $S(\omega_s,\omega_i)$ is the joint spectral intensity,
\begin{equation}
S(\omega_s,\omega_i)=|f(\omega_s,\omega_i)|^2 \mathcal{A}_s (\omega_s) \mathcal{A}_i (\omega_i=\omega_p-\omega_s),
\end{equation}
 with the assumption of an infinite number of round trips of the photons inside the cavity. The quantity $f(\omega_s,\omega_i)$ represents the joint spectral amplitude in the absence of the cavity and, in the presence of a monochromatic pump, which is the case when the crystal is pumped in a CW mode, reduces to the phase-matching function. $\mathcal{A}_j (\omega_j)$ are the Airy functions for the signal and idler photon, that, as functions of the wavelength, are expressed as
\begin{equation}
\mathcal{A}_j (\lambda_j)=\left[1+\frac{4\sqrt{R_1 R_2}}{(1-\sqrt{R_1 R_2})^2} \sin^2(\phi_j)\right]^{-1},
\end{equation}
where $\phi_j=2\pi L n_{eff}(\lambda_j)/\lambda_j$. $n_{eff}$ is the wavelength-dependent effective refractive index inside the waveguide, taking into account the material and the waveguide dispersion.

To demonstrate that the clustering effect arises from the dispersion, we use the previous equations to reproduce the measured frequency spectra previously reported~\cite{Pomarico09}. We consider a PPLN waveguide with length $L$ = 3.6\,cm, poling period  $\Gamma = 16.6$\,$\mu$m, at a temperature $T=128.6^{\circ}$\,C with mirror coating reflectivities of $R_1=R_2=0.85$. For a pump at 780\,nm, the phase-matching is satisfied for the generation of photon pairs degenerate around 1560\,nm. In this first instance, as we are interested only in the clustering effect, we can neglect the phase-matching condition, which is much larger than the effects we are studying. To model the emission spectrum we multiply the Airy function transmissions for the idler and the signal photons, whose frequencies are set by the energy conservation. The wavelength-dependence of the refractive indices for the PPLN are given by the Sellmeier equation reported in~\cite{Jundt1997}. Notice that the parameters of the Sellmeier equation have been adapted to take into account the waveguide dispersion.

\begin{figure}[t]
\begin{center}
\includegraphics[width=0.45\textwidth]{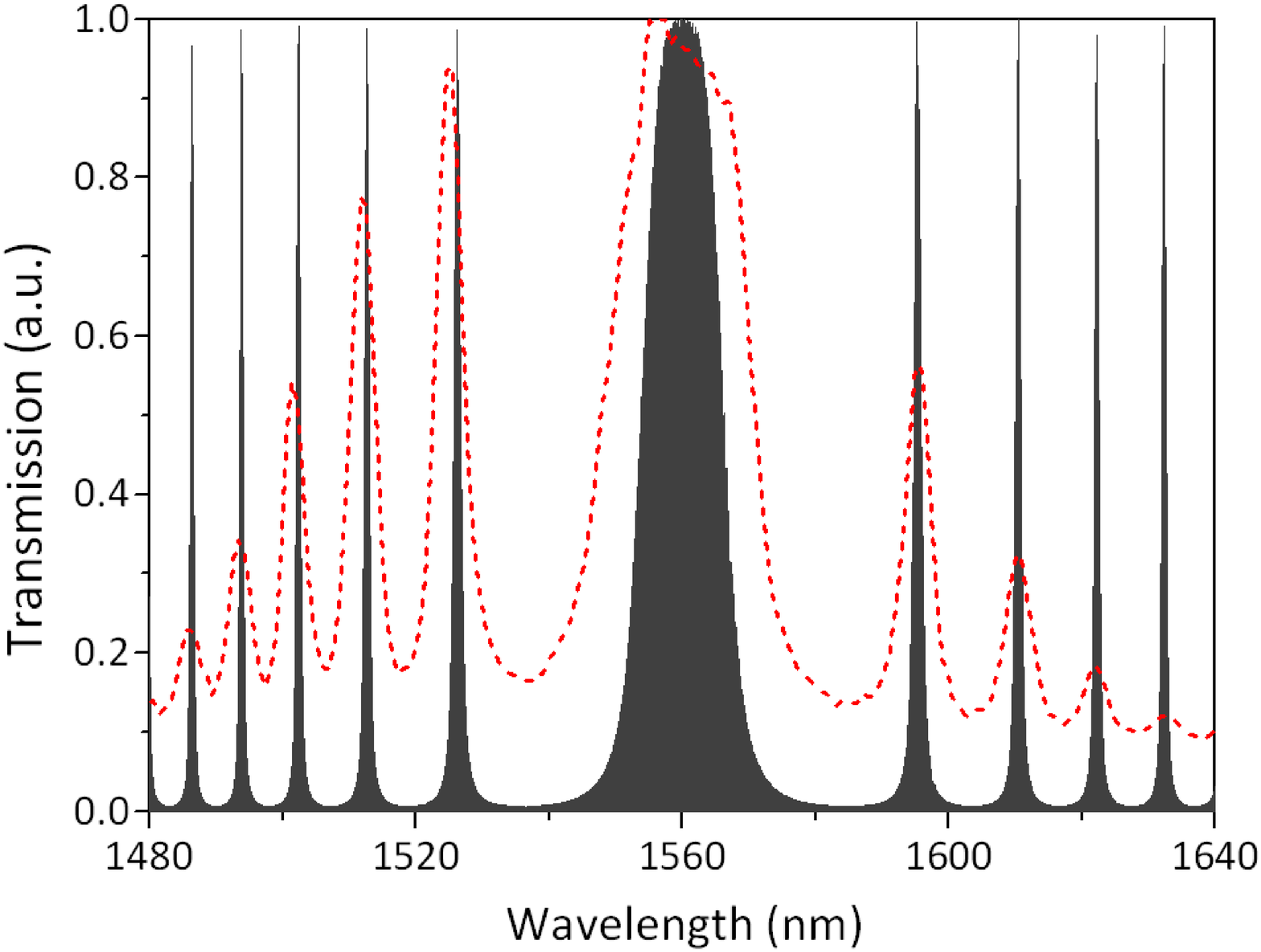}
\includegraphics[width=0.45\textwidth]{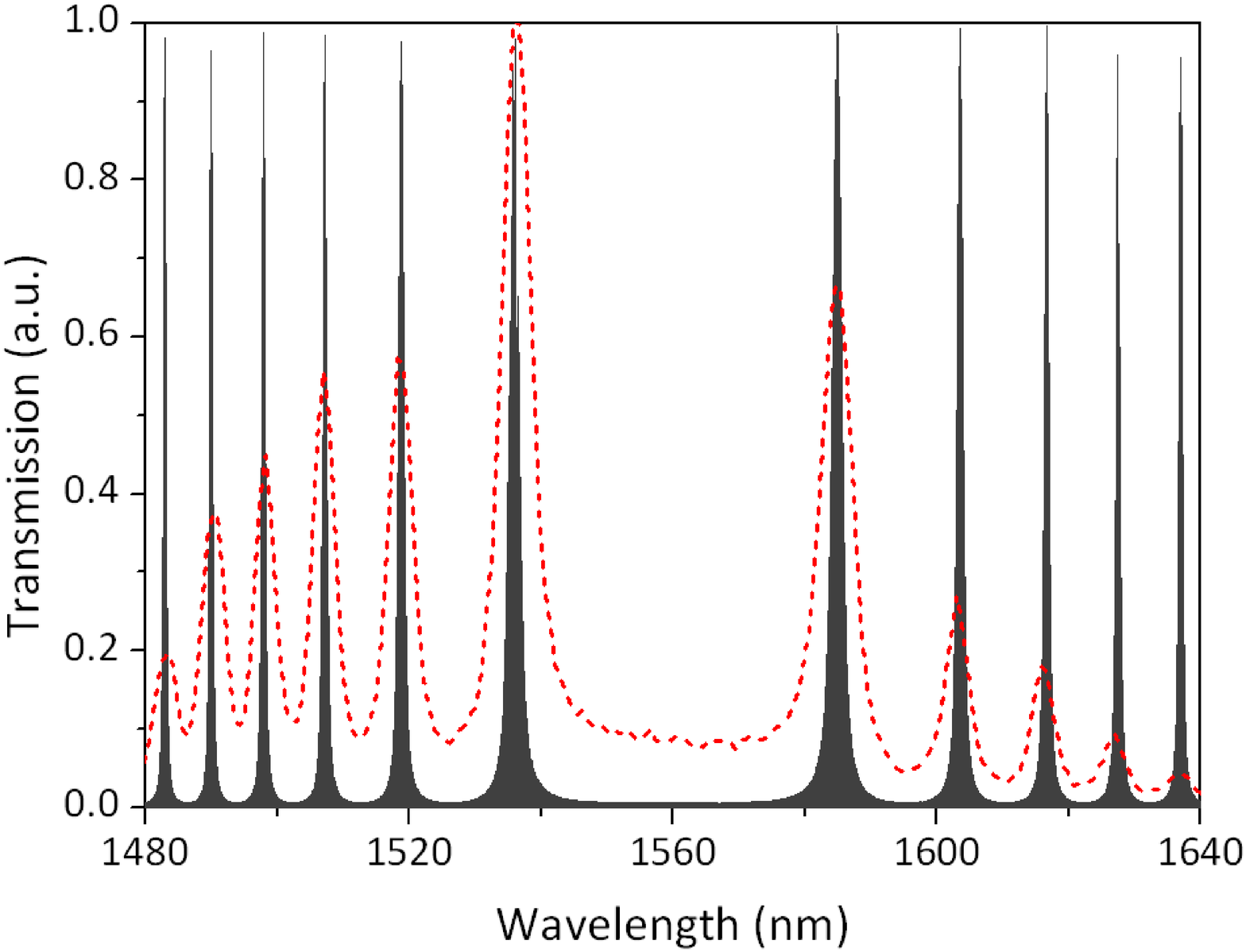}
\caption{Measured (red dashed) and calculated (black solid) SPDC spectrum for a PPLN waveguide at  two different temperatures. See text for details. }\label{fig:clustering}
\end{center}
\end{figure}
Figure~\ref{fig:clustering} shows the SPDC spectra emitted by this waveguide resonator for two different temperatures, as measured in~\cite{Pomarico09} (red dashed line) and calculated (black). The calculated clusters have a uniform height as the phase-matching and experimental detection losses have not been taken into account. The measurements were carried out with a low-resolution monochromator,
thus averaging over the fine structure within the clusters. Nonetheless, our model is in good agreement with the measured data, with the clusters found at the same frequencies and the temperature difference between the configuration with (figure~\ref{fig:clustering}a) and without (figure~\ref{fig:clustering}b) the central cluster at 1560\,nm is $\approx 0.07^{\circ}$\,C in the experiment and of 0.08$^{\circ}$\,C in the calculations. Because of the low spectral resolution of the monochromator, it is not easy to compare the measured and calculated clusters' bandwidth, but the model confirms its dependence on the finesse of the cavity, as observed in~\cite{Pomarico09}.

\section{Designing a single mode SPDC source with an integrated cavity-waveguide}
Engineering a single mode SPDC source via the clustering effect requires the clusters
to be sufficiently narrow such that they can contain only one spectral mode of the cavity and
that the FSR is sufficiently large so as to allow for easy and low-loss extraction of the desired
mode.

\subsection{Determining the finesse}

A cluster contains $M$ modes when the difference between the spectral range corresponding to these modes for the idler and the same range for signal is equal to the single mode bandwidth.
In quantitative terms, the number of modes within the cluster bandwidth can be expressed as:
\begin{equation}
M (FSR_i - FSR_s)=\Delta\nu,
\end{equation}
where $FSR_{i,s}$ is the free spectral range for the idler and signal and $\Delta\nu$ is the mode bandwidth.
We consider for $\Delta\nu$ the mean value of the idler and signal bandwidths:
\begin{equation}
\Delta\nu=\frac{\Delta\nu_i+\Delta\nu_s}{2}=\frac{FSR_i+FSR_s}{2\mathcal{F}},
\end{equation}
where $\mathcal{F}$ is the finesse of the cavity, assumed to be the same for the idler and signal photons. Since $FSR_{i,s}=c/(2 N_{i,s} L)$, where $N_{i,s}$ is the group index at the idler and signal wavelength, we can solve for the number of modes inside the cluster and determine the constraints on the finesse to have just a single mode:
\begin{equation}
M=\frac{1}{2\mathcal{F}}\Big(\frac{N_s+N_i}{N_s-N_i}\Big) \hspace{1cm} \Longrightarrow \hspace{1cm} \mathcal{F}_{M=1}=\frac{1}{2}\Big(\frac{N_s+N_i}{N_s-N_i}\Big).
\end{equation}\label{eq:finesseN1}
\begin{figure}[h!]
\begin{center}
\includegraphics[width=0.48\textwidth]{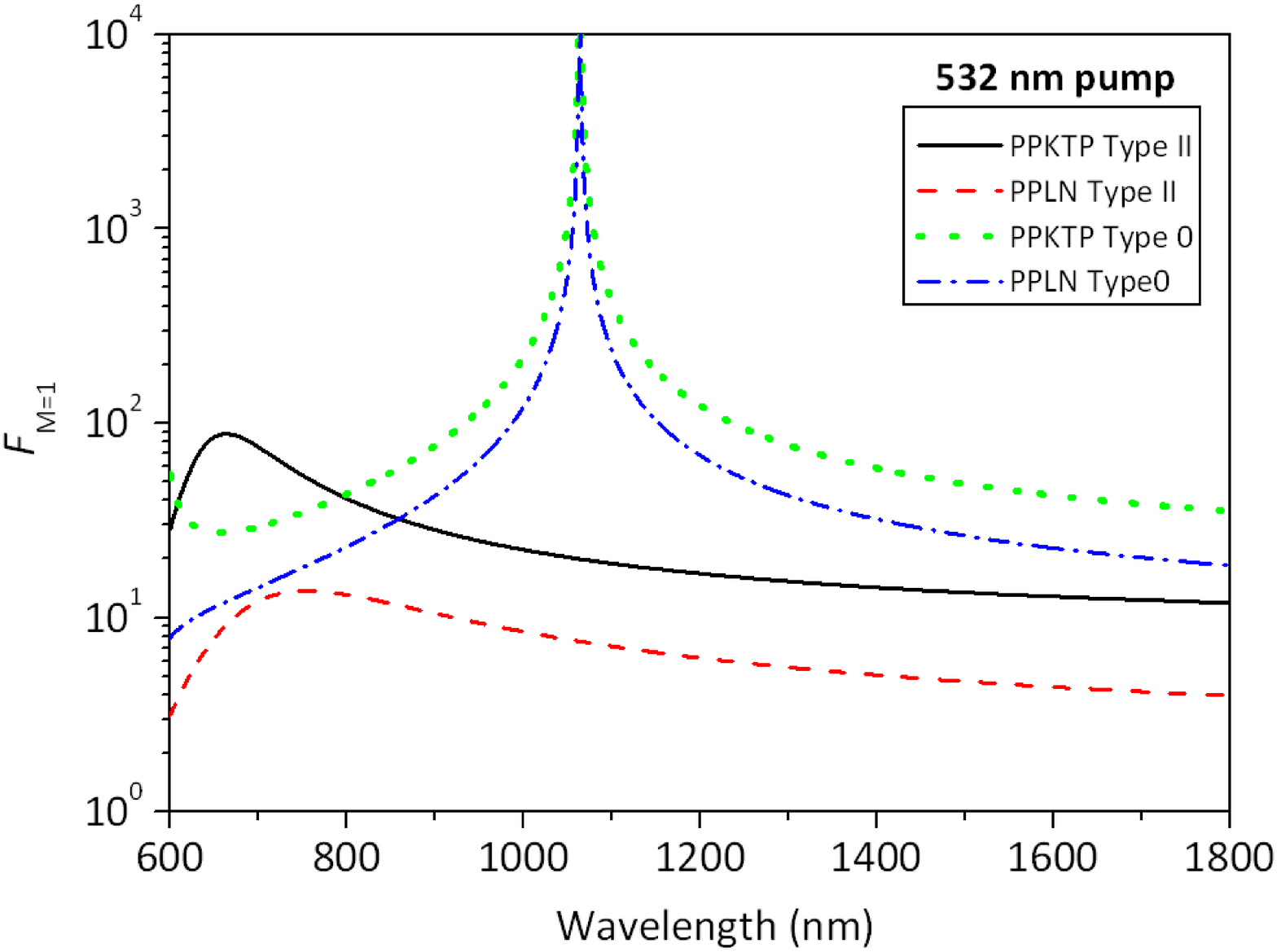}
\includegraphics[width=0.48\textwidth]{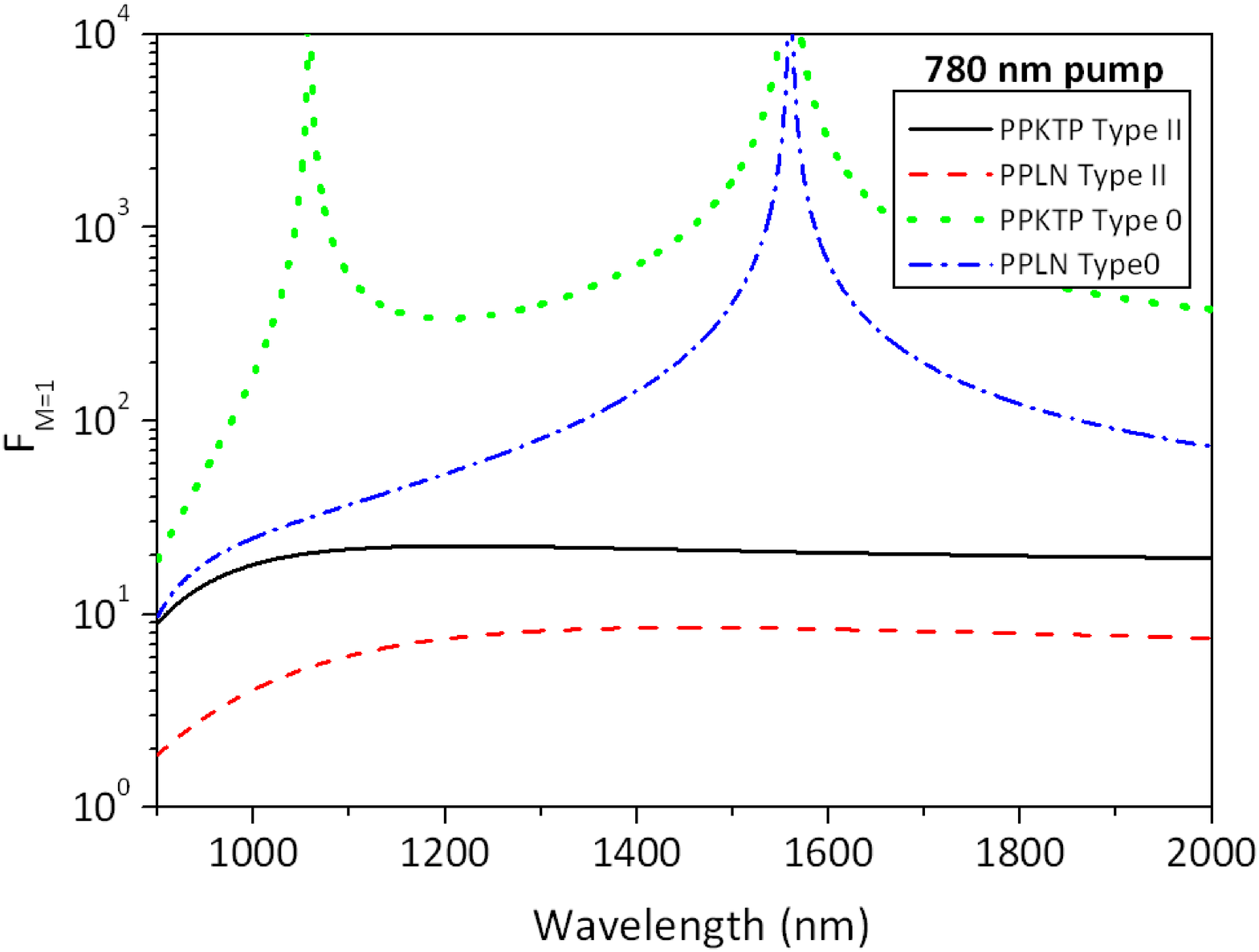}
\caption{The finesse $\mathcal{F}_{M=1}$, corresponding to one mode in the clusters, for a 532\,nm (left) and 780\,nm (right) pump, as a function of the signal wavelength. Type 0 (eee) and Type II (eoe) interactions in PPKTP and PPLN are considered.}\label{fig:finesse}
\end{center}
\end{figure}
Therefore, finesse values larger than $\mathcal{F}_{M=1}$ are good to select single spectral modes.
In figure~\ref{fig:finesse} we give examples for the finesse $\mathcal{F}_{M=1}$, as a function of the signal wavelength, in the case of a 532\,nm or 780\,nm pump. The different dispersion in the case of Type~0 (eee) and Type~II (eoe) interactions for PPLN and PPKTP are considered. For the dispersion of the ordinary refractive index in bulk PPLN and PPKTP we use the Sellmeier equations reported in~\cite{Edwards1984} and~\cite{Bierlein1989} respectively. The curves in figure~\ref{fig:finesse} depend only on the energy conservation and the dispersion inside the crystals, and do not take into account the phase-matching condition. Phase-matching at different wavelengths can be obtained with specific values of the poling period.

It is evident from this result that Type II PPLN and PPKTP crystals require lower values of the finesse in the telecom regime, making the realization of integrated cavity-waveguides easier. Notice that these considerations only provide a lower bound for the value of the finesse and, according to the needs, one has to define the length and the reflectivities of the mirrors of the cavity in order to have the desired mode bandwidth and internal losses.

\subsection{Finding the optimal parameters of the integrated cavity-waveguide}

In order to design a single-mode source with an integrated waveguide resonator with a precise value
of the finesse, one has to know the absorption coefficient and optimize the values of $R_2$ and $L$
to take into consideration the losses inside the crystal. Therefore, let us calculate the probability $p_{out}$ of emitting photon pairs. With a reasoning similar to that present in~\cite{Pomarico09}, assuming the initial generation of the photon pairs in the center of the waveguide, we find
\begin{eqnarray}\label{eq:pout}
p_{out} &=& 10^{-(\alpha /2)L/10}(1-R_2) \sum_{n=0}^{\infty} \Big(R_1R_2 10^{-4\alpha L/10}\Big)^n \\
&=& \frac{10^{-(\alpha /2)L/10} (1-R_2)}{1- R_1 R_2 10^{-2\alpha L/10}}.\nonumber
\end{eqnarray}
The probability $p_{out}$ depends on the length $L$ of the cavity-waveguide. Clearly, for larger values of $L$ the probability $p_{out}$ decreases, though, to keep the finesse fixed, the reflectivity $R_2$ needs to be increased. Therefore, shorter crystals guarantee a small amount of overall loss and a large free spectral range, which is necessary for the efficient selection of a single spectral mode in the clusters.

\subsection{Purity of the photon state}
The purity of the state produced by a SPDC source depends on the quality of the spatial and spectro-temporal mode emission. A single spatial mode can be selected with the use of a single mode optical fibre at the cost of some losses that depend on the emission properties of the source. However, single-mode waveguides have been shown to have very efficient ($\sim90\%$) coupling to single-mode fibers~\cite{Roussev04a}.

The spectro-temporal purity of the state produced by integrated OPO devices depends on the number of modes in which the photons are emitted. In order to achieve high purity, one single spectro-temporal mode with bandwidth $\Delta\nu$ has to be isolated. This can be done using the clustering effect described above. Then, the mode has to be resolved \cite{Halder2007,Huang2010} by adopting a detector with a time jitter smaller than the coherence time of the photons $\tau_c=1/(\pi\,\Delta\nu)$ \cite{Lu2000}. Thus, having fast detectors  ensures a single mode operation. Standard detectors have a timing resolution in the order of hundreds of picoseconds, requiring $\Delta\nu\ll1\,GHz$, which is easily satisfied by this source.

\section{Optimized single mode integrated OPO source}
Let us now consider a specific device design. A CW laser at 780\,nm pumps a Type II PPLN cavity-waveguide. From equation~\ref{eq:finesseN1} and using the Sellmeier equation reported in~\cite{Jundt1997}, we know that, in order to obtain a single mode source at 1560\,nm, we need a finesse larger than $10$. Now, to have a sufficiently large $FSR$, we consider a crystal of length $L$=1\,mm. By assuming $R_1\approx$1 and $R_2\,=\,0.95$, which are attainable with current reflection coating technology, we obtain a finesse of 116.2 from equation (\ref{eq:finesse}). For the absorption coefficient we use the value of $\alpha$=0.06\,dB/cm, indicated in~\cite{Pomarico09}. Using the equation (\ref{eq:pout}), we find a value of $p_{out}$ of 95\%.

\begin{figure}[!h]
\begin{center}
\includegraphics[width=0.6\textwidth]{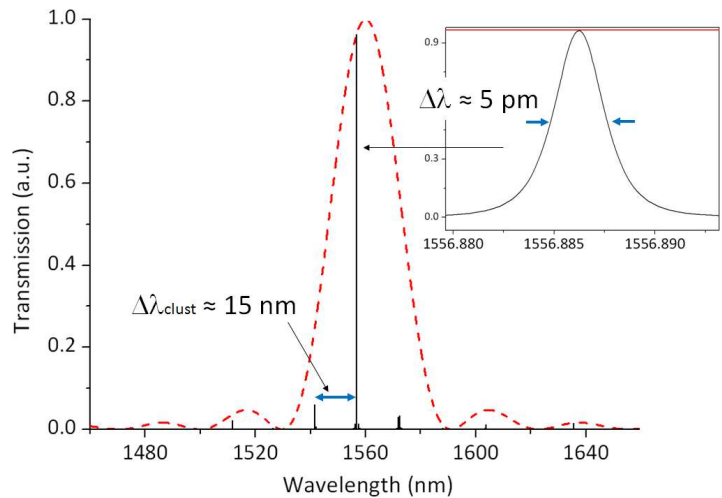}
\caption{Frequency spectrum, affected by the clustering effect, of an integrated cavity-waveguide whose parameters are indicated in the text. The red (dashed) line represents the phase-matching envelope. Right inset: zoom on the central single spectral mode. }\label{fig:spectrum}
\end{center}
\end{figure}
By adopting the model described in section \ref{par:clustering}, we calculate the resonances of the cavity affected by the clustering effect at the temperature of $T=80.14^{\circ}$C. This is illustrated in figure~\ref{fig:spectrum}, where the red (dashed) curve represents the phase-matching envelope, calculated using the model reported in~\cite{Fiorentino2007}. The phase mismatch $\Delta k$ can be expressed as $\Delta k = k_p - k_i - k_s - 2\pi/\Lambda$, where $k_{p,i,s}$ are the projection of the pump, idler and signal wave-numbers to the direction of the waveguide and $\Lambda$ is the effective poling period, which ensures the phase-matching.

As one can see from figure~\ref{fig:spectrum}, the spectrum is characterized by a series of resonances, which are inside distinct clusters. The phase-matching envelope contains only three single mode clusters, which are present at a spectral distance of $\Delta \lambda_{clust}\approx 15$\,nm: it is therefore possible to filter the central mode, for example, by simply using a prism or a notch filter, which have negligible transmission loss. This device represents an example of a single-mode SPDC cavity-waveguide which does not need additional, lossy, narrowband filters and produces photons with a bandwidth of $\Delta \lambda = 4.6$\,pm (right inset)

\section{Discussion}
Engineering spectrally pure photon states, avoiding the use of external filters, is fundamental for performing a variety of quantum communication applications and complex multi-photon experiments.
Because of the gaussian loss induced by the filters, some protocols, such as Device Independent QKD, simply cannot be implemented, since it becomes impossible to close the detection loophole. On the other hand, for the same gaussian filtering factor, multiphoton experiments becomes extremely slow and practically unfeasible, because the overall amount of loss scales exponentially with the number of photons. 
\begin{figure}[!h]
\begin{center}
\includegraphics[width=0.6\textwidth]{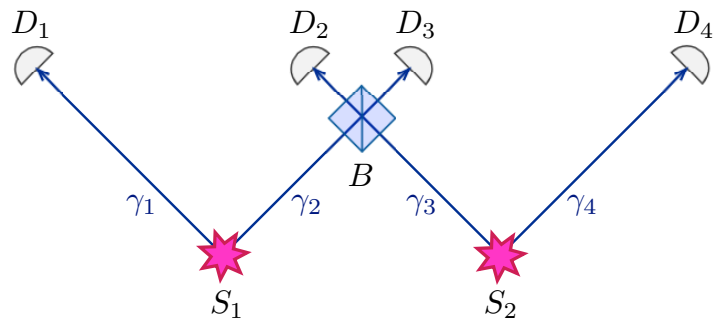}
\caption{Photon interference scheme. The visibility of the Hong-Ou-Mandel interference of the photons $\gamma_2$ and $\gamma_3$ is maximum when these two photons are in pure and indistinguishable states. In an entanglement swapping experiment, photons $\gamma_2$ and $\gamma_3$ are entangled with their respective counterparts $\gamma_1$ and $\gamma_4$. When the entanglement is in energy-time, photons $\gamma_2$ and $\gamma_3$ still need to be spectrally single mode in order to achieve the maximum interference visibility.}
\label{fig:HOM}
\end{center}
\end{figure}

Let us consider the interference scheme illustrated in figure~\ref{fig:HOM}. We can see this as the Hong-Ou-Mandel interference between the photons $\gamma_2$ and $\gamma_3$, heralded by the detectors $D_1$ and $D_4$ respectively. Achieving unit interference visibility at beamsplitter $B$ requires that the photons are in pure and indistinguishable states.

Now, we can increase the complexity of the system through entanglement swapping operations. In the experiment reported in \cite{Halder2008}, an initial coincidence detection after the beamsplitter performed a Bell state measurement (BSM) that was used to herald the generation of time-bin entanglement of the photons $\gamma_1$ and $\gamma_4$. This particular experiment had $\sim3\,$dB transmission loss, as well as the Gaussian loss, on each photon, due to an external filtering system. Using the approach proposed here, we would be able to directly engineer the photons in the cavity and, as the photons are spectrally pure and indistinguishable, we would only be subject to the loss associated with spatial coupling from the waveguide into the fiber, which can in principle be around 90\%~\cite{Roussev04a}. In this previous experiment the visibility was also limited due to the external filters moving with respect to each other and the coherence time of the photons not being sufficiently large compared to the detector jitter. In our proposed system these issues should be resolved and we could expect near unit interference visibility and an increase in the 4-fold coincidence rates by almost two orders of magnitude, before considering the improvements in fiber coupling and detector efficiencies achieved over the last few years.

\section{Conclusions}
We have outlined a model for the clustering effect in an integrated cavity-waveguide SPDC photon pair
source. We have shown how this effect can be used for the
engineering of integrated devices, which, in conjunction with fast photon detectors, can efficiently
generate pure photonic states with current technologies. Due to the very narrow
bandwidth of the photons, we expect that this will be of great interest, not only for quantum
communication, but for a wide range of emerging quantum technologies and more fundamental
investigations such as all-optical, loop-hole-free, Bell experiments.

\section*{Acknowledgment}
We thank H.~Zbinden, N.~Sangouard and N.~Gisin for valuable discussions. This work was supported by the EU projects Qessence, QuReP and the Swiss NCCR-QSIT.

\end{document}